\documentstyle[12pt]{article}
\begin{document}
\def\dag{\dagger}
\def\del{\partial}
\def\ybh{\hat{\bar\y}}
\def\dba{\dot{\bar A}}
\def\a{\alpha}     \def\A{A}        
\def\b{\beta}      \def\B{B}        
\def\g{\gamma}     \def\G{\Gamma}   
\def\d{\delta}     \def\D{\Delta}   
\def\e{\epsilon}   \def\E{E}        
\def\z{\zeta}      \def\Z{Z}        
\def\j{\eta}       \def\J{H}        
\def\q{\theta}     \def\Q{\Theta}   
\def\i{\iota}      \def\I{I}        
\def\k{\kappa}     \def\K{K}
\def\l{\lambda}    \def\L{\Lambda}  
\def\m{\mu}	   \def\M{M}        
\def\n{\nu}        \def\N{N}        
\def\x{\xi}        \def\X{\Xi}      
\def\o{o}          \def\O{O}        
\def\p{\pi}        \def\P{\Pi}     
\def\r{\rho}       \def\R{P}        
\def\s{\sigma}     \def\S{\Sigma}   
\def\t{\tau}       \def\T{T}        
\def\u{\upsilon}   \def\U{Y}        
\def\f{\phi}       \def\F{{\mit\Phi}}     
\def\h{\chi}
\def\y{\psi}       \def\H{X}        
\def\ps{\y}
\def\yd{\y^{\dag}}
\def\yb{\bar{\y}}
\def\yh{\hat{\y}}
\def\yhd{\hat{\y}^{\dag}} 
\def\Y{{\mit\Psi}}      
\def\Ps{\Y}
\def\w{\omega}     
\def\W{\Omega}   
\def\br{\langle}
\def\ke{\rangle}
\def\ve{\vert}
\def\inf{\infty}
\def\Winf{$W_{\infty}\  $}
\def\winf{$w_{\infty}\  $}
\def\to{\rightarrow}
\def\kvecz{\ve z_1 z_2 \cdots z_n\ke}
\def\bvecz{\br z_1 z_2 \cdots z_n \ve}
\def\kveczp{\ve z' _1 z' _2 \cdots z_n\ke}
\def\kveczpp{\ve z" _1 z" _2 \cdots z_n\ke}
\def\bveczp{\br z' _1 z' _2 \cdots z_n \ve}
\def\bveczpp{\br z" _1 z" _2 \cdots z_n \ve}
\def\zbar{\bar{z}}
\def\zintm{d^2 z e^{-|z|^2}}
\def\zintmp{d^2 z' e^{-|z'|^2}}
\def\to{\rightarrow}
\def\tr{{\rm tr}}
\font\ti=cmti10 scaled \magstep2
\def\ba{{\cal A}}
\def\bF{{\cal F}}
\def\bE{{\cal E}}
\def\bb{b}
\def\adel{{\hbox{$\> \buildrel {\leftrightarrow}\over {\partial}\>$}}}
\def\vf{\varphi}
\begin{titlepage}
 \renewcommand{\thefootnote}{\fnsymbol{footnote}}
    \begin{normalsize}
  \begin{flushright}
                IASSNS-HEP-97/21   \\
                CCNY-HEP 97/3 \\
                 March 1997 \\
                 
     \end{flushright}
    \end{normalsize}
    \begin{Large}
       \vspace{1cm}
       \begin{center}
         {\bf  Breakdown of Universality \\ in Random Matrix Models
         } \\
       \end{center}
    \end{Large}

  \vspace{10mm}
\begin{center}
  S. Iso\footnote{On leave of absence from National Laboratory
  for High Energy Physics (KEK) \\
   \hspace*{5mm} iso@theory.kek.jp} \\ 
\vspace{5mm}   
 {\it Institute for Advanced Study}\\
 {\it Princeton, New Jersey  08540}  \\ 
\vspace{5mm} 
{ \it and } \\ 
\vspace{5mm}
  An. Kavalov\footnote{kavalov@scisun.sci.ccny.cuny.edu} \ \ 
\\
\vspace{5mm} 
{\it Physics Department, City College 
          of the City University of New York}\\
{\it New York, NY 10031}  \\
\end{center}
\vspace{15mm}
\begin{abstract}
\noindent 
We calculate  smoothed correlators for a large random matrix model
with a potential containing products of two traces $\tr W_1(M) \cdot
\tr W_2(M)$ in addition to a single trace $\tr V(M)$. 
Connected correlation function of density eigenvalues 
receives  corrections besides the universal part derived by
Br\'{e}zin and Zee
and it is no longer universal in a strong sense.
\end{abstract}
\end{titlepage}
\vfill
\eject
\newpage
\section{Introduction}
Random matrix theory was originally introduced by Wigner and studied
in detail by Dyson and Metha to investigate  statistical
properties of energy levels of heavy nuclei \cite{Metha}.
It has been applied to various fields recently such as
quantum chaos \cite{Bohigas90}, 
quantum dots or 2d discretized gravity \cite{Gravity95}.
It has also been applied to quantum transport problem of mesoscopic
wires reviewed in \cite{Bee96}.
\par
In the above mentioned contexts universal behavior of correlation functions
of eigenvalues has been discussed. 
There are two types of universalities, 
one for a short distance behavior (where correlation functions
oscillate rapidly) \cite{BHZ96}
and the other for a smoothed correlator of distance scale larger
than the rapid oscillation \cite{BZ93a}. In this letter 
we study the latter universal
behavior of correlation functions. 
\par
The universal large scale behavior of density correlation 
functions was pointed out by Br\'{e}zin and Zee who stressed its 
importance in disordered systems  though it
was already derived  by Ambjorn, Jurkiewicz and Makeenko \cite{AJM90} 
in the context of 2d discretized gravity.
They considered a random matrix theory for large  matrices $M$, 
whose statistical weight is given by
\begin{equation}
{1 \over Z} e^{-N \tr V(M)}
\label{int10}
\end{equation}
and showed that,  in the large matrix limit, the 
 connected  two point correlation function 
of density of eigenvalues $\langle \rho(x) \rho(y) \rangle_{c}$
has a universal form which has no
explicit dependence on the potential $V$.
For a symmetric potential it is
\begin{equation}
- {1 \over 2 \pi^2 \lambda (x-y)^2 }{a^2 -xy \over \sqrt{(a^2-x^2)(a^2 - y^2)}}
\label{int20}
\end{equation}
where $a$ and $-a$ are the end points of the density distribution and 
$\lambda = 1/2, 1, 2$ corresponding to orthogonal, unitary or 
symplectic ensembles.
This universal form has been  calculated in various methods 
\cite{AJM90, BZ93a, Bee93, Forr95, IMS}.
and its $1/N$ corrections were
also studied for general ensembles \cite{AJM90, ACKM, Itoi}.
\par
Physical implication of this universality to the universal conductance 
fluctuation of mesoscopic wire is discussed in a recent review
by Beenaker \cite{Bee96}.
Universality here means that fluctuation of conductance 
$\delta G$ in a metalic regime is of order
$e^2/ h$, independent of sample size or disorder strength.
This strong suppression of conductance fluctuation 
was first pointed out by Altshuler \cite{Alt} and Lee and Stone
\cite{Lee}.
Also, when magnetic field is applied, variance $(\delta G)^2$
of the conductance  decreases exactly by a factor of two
compared to the fluctuation without magnetic field.
The system can be modeled by random matrix theory for transmission
matrix \cite{Bee96}.
Since the  variance of conductance is proportional to the connected
density-density correlation function,
the above universality of conductance fluctuation 
is mapped to the universality of correlation functions in random 
matrix theory. Looking back to the equation (\ref{int20}),
first it is $O(N^0)$ and independent of the system size ($N$) or 
details of disorder ($V$). Next it decreases exactly by a factor
of two when magnetic field is applied ($\lambda$ is changed from 
$1/2$ to $1$ by applying magnetic field.)
It is quite nice that such a simple model as the random matrix
theory can explain some of important 
features of complicated systems.
\par
Since the main guiding principle of random matrix theory is 
randomness and symmetry, it is natural to ask whether this 
universality still holds for more general ensemble which 
is invariant under symmetry rotation.
Simple generalization of the potential (\ref{int10}) is to add
products of traces to the statistical weight of matrices;  
\begin{equation}
{1 \over Z} e^{-N \tr V(M) - \tr W_1 (M) \cdot \tr W_2(M)}.
\end{equation}
This ensemble is invariant under $M \rightarrow UMU^{-1}$ where
$U$ is an orthogonal, unitary or symplectic matrix correspondingly.
The ensemble with this  generalized potential was studied by 
\cite{DDSW} in the context of 2d gravity.
Universality in this ensemble was discussed by \cite{BZ93, KHKZ94}.
Br\'{e}zin and Zee \cite{BZ93}
 argued that this model is equivalently described by an ensemble
with an effective single trace  potential $V_{eff}$
and concluded that
the universality still holds.
\par
In this letter we study this generalized ensemble and
obtain density of eigenvalues and its correlation functions explicitly.
In section 2, we review a collective field theory approach to
an ensemble with a single trace potential and show
how the universal behavior of correlation functions emerges.
In section 3, we generalize it to an ensemble with a multi trace 
potential (containing products of two traces).
We show that the universality is broken and the correlation function
is no longer universal in the strong sense.
Finally in discussion, we discuss 
why the argument by Br\'{e}zin and Zee \cite{BZ93} 
does not hold for correlation functions.
In Appendix A and B, we prove useful formulas and in Appendix C
we give an example of a correlation function for a simple ensemble
with a multi trace potential.
\section{Single Trace Matrix Model}
In this section  we review how to calculate density of 
eigenvalues and its correlations in random matrix theory
in the large $N$ limit($N$ is a size of matrices)
and  show how the universal form  of a 
two-point  correlation function
emerges.
We consider a matrix model with
an ordinary  single trace potential in the Collective Field 
approach \cite{JS}. This approach is easily  
generalized to a multi trace potential, discussed
in the next section.     
\par
The free energy $F[V]$ is defined as follows:
\begin{equation}
e^{- N^2 F[V]} \equiv \int {dM \over Vol} 
\  e^{-N \tr V(M)} = 
\int d x_1 ... d x_N \ \D^{2 \l}  \   
e^{-N \sum_{i=1}^{N}V(x_i)} 
\label{a10} 
\end{equation}
where $Vol$ is the volume of gauge symmetry group,
$\D = \prod_{1 \le i < j \le N} (x_i - x_j)$ is
the Van der Monde determinant, and $\l = 1/2, 1$ or $2$
for orthogonal, unitary or symplectic ensembles correspondingly.
Partition function is invariant under orthogonal,
unitary or symplectic rotations and the matrix integral can be
reduced to integrals over its eigenvalues $x_i$ ($i=1 \sim N$).
(Normalized) density of eigenvalues is defined by
\begin{equation}
\r(x) = {1 \over N} \sum_{i=1}^N \d(x - x_i).
\label{a12} 
\end{equation}
Connected density-density correlation functions can be obtained
from the free energy by taking variational derivatives 
with respect to $V(x)$:
\begin{equation}
\br \r(x_1) \r(x_2) ... \r(x_m) \ke_c = 
{(-)^{m-1}  \over N^{2m-2}} 
{\d \over {\d V(x_1)}} 
{\d \over {\d V(x_2)}} \ ...  \    
{\d \over {\d V(x_m)}} \  F[V]  .
\label{a11}
\end{equation}
It is obvious now that 
the leading non-vanishing term for an 
$m$-point (unnormalized) density correlation function 
is $O(N^{2-m})$.
The standard procedure of collective 
field theory is to rewrite  integrals
over eigenvalues in terms of a functional integral over  density 
$\rho(x)$.
Inserting $one$ to the equation (\ref{a10})   
\begin{equation}
1 = \int D \r(x) \prod_x \d \bigg( \r(x) - 
{ 1 \over N} \sum_{i=1}^N \d(x - x_i) \bigg) ,
\end{equation}
we obtain up to an overall constant (see Appendix A)
\begin{equation}
e^{ - N^2 F[V]} = \int D\r(x) d\s \  
e^{N^2 \l \int dx dy \r(x) ln \ve x - y \ve \r(y) - 
N^2 \int dx \r(x) V(x) + 
i \s \big( \int dx \r(x) - 1  \big) } \  J[\r] .
\label{a14} 
\end{equation}
 Lagrange multiplier $\sigma$ is introduced to impose 
the constraint that an integral of $\rho(x)$ is normalized to one.
We are interested in the large $N$ behavior
and the Jacobian $J[\r]$ can be neglected in this limit
(see \cite{KS} for details).  
The resulting integral over $\r$ can be evaluated
by steepest descent method which requires to solve the following
equations of motions \cite{JS}: 
\begin{eqnarray}
0 & = & 2 N^2 \l \int d\x \ ln \ve x-\x \ve  \r_0(\x)  
- N^2 V(x) + i \s_0, 
\label{a19}    \\ 
0 & = & \int dx \ \r_0(x) - 1  .
\label{a20}
\end{eqnarray}
By differentiating  the first equation we have
the equation of BIPZ
\cite{BIPZ}
\begin{equation}
P \int d\x \  {\r_0(\x) \over {x - \x}} = {V^{'}(x) \over {2\l}}
\label{a25}
\end{equation} 
and it determines the stationary value $\r_0(x)$.   
We assume here that $\r_0(x)$ is equal to $zero$ 
for $x < a$ or $x > b$ ({\it one-cut} from $a$ to $b$). Then the
solution of this Cauchy integral equation is given by     
\cite{MI}:   
\begin{equation}
\r_0(x) = {1 \over \pi} {1 \over \sqrt{(x-a)(b-x)}}
\bigg\{ {1 \over {2 \pi \l}} P \int_a^b  d\x \ 
{\sqrt{(\x-a)(b-\x)} \over \x - x} V^{'}(\x)\ +\ 1 \bigg\} 
\label{a30}
\end{equation}
The second term in the curly bracket is a solution of a homogeneous
integral equation (i.e \ set  the r.h.s. of eq.
(\ref{a25}) $zero$)
and its coefficient is determined such that 
$\r_0(x)$ satisfies equation (\ref{a20}), while 
the first term does not contribute to the integral of eq.(\ref{a20})
(see Appendix B).
\par
Equation (\ref{a19}) defines $\s_0[V]$ as a functional of
$V$ and, of course,  independent of $x$ even though it enters to the 
solution manifestly. Unknowns left are the positions
of the end points of the cut $[a,b]$. 
By choosing boundary conditions $\r_0(a) = 0$ and
$\r_0(b) = 0$ we get:
\begin{eqnarray} 
\r_0(a) & = & 0  \ \ \ \ \ \ \Longleftrightarrow \ \ \ \ \ \ 
{1 \over {2\pi \l}} \int_a^b d\x \   
\sqrt{b-\x\over{\x-a}} \  V^{'}(\x)    
\ + \ 1 = 0,     
\label{a39}  \\ 
\r_0(b) & = & 0  \ \ \ \ \ \ \Longleftrightarrow \ \ \ \ \ \ 
 - {1 \over {2\pi \l}} \int_a^b d\x   \   
\sqrt{\x-a \over{b-\x}} \  V^{'}(\x)  
\ + \ 1 = 0. 
\label{a40} 
\end{eqnarray}
Equations (\ref{a39}) and (\ref{a40}) determine  $a = a[V]$ and
$b = b[V]$ as functionals of $V(x)$, and should be solved
first. This completes the solution  $\r_0(x)$ through the  
equation (\ref{a30}). For a simple case $V(x) = x^2 /2 $  
this procedure reproduces the famous Wigner 
semi-circle solution. 
\par
Dependence of $\rho_0(x)$ on $V(x)$ comes both 
explicitly, and implicitly through the end points
and in order to find connected correlation functions 
(\ref{a11})      
we have to know $V$-dependence of the end points.
However we have an important relation that,
given the boundary conditions 
(\ref{a39}) and (\ref{a40}), the solution (\ref{a30}) 
 satisfies 
\begin{equation}
{\del \over{\del a}} \r_0(x) = 
{\del \over{\del b}} \r_0(x) = 0. 
\end{equation} 
This  can be checked by straightforward calculations. 
Therefore upon variation of $\r_0(x)$ with respect to $V(x)$ only
manifest dependence on $V(x)$ in (\ref{a30}) is relevant, 
while $V$-dependences of the boundaries $a$ and $b$ are cancelled out. 
Since the two-point correlation function is  obtained
from $\rho_0(x)$ by taking a variational derivative,   
this is essentially  statement
of the universality of a two-point correlation function:
\begin{equation}
N^2 \br \r_0(x) \r_0(y) \ke_c 
= - {\d \over{\d V(y)}} \r_0(x) 
= {1 \over{2\p^2 \l}} 
{\del \over{\del y}} \bigg\{ 
\sqrt{(y-a)(b-y)\over{(x-a)(b-x)}} {P\over{y-x}} 
\bigg\}   
\label{a50}   
\end{equation}
All dependence on the potential $V(x)$ 
is implicit only through
the end-points $a$ and $b$. Here $P$ stands for the
principal value: 
$P {1 \over{x}} = \lim_{\e\to 0} {x/(x^2 + \e^2)}$.
This result has been obtained
 in various papers by various methods 
\cite{BZ93a, AJM90, Bee93, Forr95, Itoi, IMS}.   
\par
Collecting all results together we obtain 
for the free energy $F[V]$ the following expression 
(in the $N\to\inf$ limit):  
\begin{equation} 
F[V] = \int_{a[V]}^{b[V]} dx  \   \r_0(x) 
 V(x)  - \lambda \int_{a[V]}^{b[V]} 
dx \ dy \rho_0(x) ln|x-y| \  \rho_0(y).  
\label{a60}    
\end{equation}
As expected, density of eigenvalues is given by the saddle point 
solution;
\begin{eqnarray}
&& \br \r(x) \ke_c = {\d F[V] \over{\d V(x)}}   \nonumber \\
&&= \r_0(x)  +  \int_{a[V]}^{b[V]} d \xi \  
{\d \rho_0 ( \xi ) \over{\d V(x)}} 
(V(\xi ) - 2 \lambda \int dy \ ln|x-y| \ \rho_0(y) )
\nonumber \\
&&=  \r_0(x) - {i \sigma_0 \over{N^2}}
 \int_{a[V]}^{b[V]} d\xi \  
  \langle \rho_0(\xi ) \rho_0 (x) \rangle_{c} = \r_0(x).
\end{eqnarray}
In principle $1/N$ corrections 
can be calculated by evaluating 
the Jacobian $J[\rho]$ \cite{KS} and fluctuation around 
the saddle point. 
\section{Multi Trace Matrix Model}  
\par
Let us now generalize our approach to a
multi trace case. 
We consider an ensemble with the following statistical weight 
\begin{equation}
{1 \over Z} \exp \bigg( -N \tr V(M) 
- {1 \over 2} \sum_{\a,\b = 1}^K \w_{\a\b}
\tr W_{\a}(M) \tr W_{\b}(M)  \bigg).
\end{equation}
In this case, after integrating over the 
angular variables, free energy is given by:  
\begin{equation}
e^{- N^2 F[V]} \equiv 
\int d x_1 ... d x_N \  \D^{2 \l}
\ e^{-N \sum_{i=1}^{N}V(x_i) - {1\over{2}}
\sum_{\a,\b = 1}^K \w_{\a\b}
\sum_{i,j=1}^N W_{\a}(x_i) W_{\b}(x_j)  }    
\label{b10}
\end{equation}
and as before connected $m$-point correlation functions can
be obtained by taking functional derivatives 
with respect to $V(x)$ (see (\ref{a11})).    
Expressing $F[V]$ in terms of the density $\r(x)$
we obtain:  
\begin{eqnarray}
e^{ - N^2 F[V]} = \int D\r(x) d\s \  
e^{N^2 \l \int dx dy \r(x) ln \ve x - y \ve \r(y) - 
N^2 \int dx \r(x) V(x)}  \times \nonumber \\  
\times  e^{- N^2 \sum_{\a,\b = 1}^K \w_{\a\b} 
\int dx \ \r(x) W_{\a}(x)  \int dy \ \r(y) W_{\b}(y)  
+ i \s \big( \int dx \r(x) - 1  \big)}J[\r]
\end{eqnarray}
with the same 
Jacobian $J[\r]$ as in the previous section. 
Again in the leading
order in $N$ we can set $J[\r] = 1$. 
The steepest descent equations are: 
\begin{eqnarray}
&0& =2 N^2 \l \int d\x ln \ve x-\x \ve  \r_0(\x)  
- N^2 V(x) - N^2 \sum_{\a,\b =1}^K \w_{\a\b} W_{\a}(x) c_{\b}  
+ i \s_0, 
 \nonumber \\
\label{b19}    \\ 
&0& =\int dx \ \r_0(x) - 1   
\label{b20}
\end{eqnarray}
where $c_{\beta}$'s  are constants 
\begin{equation}
c_{\b} \equiv \int dx \ \r_0(x) W_{\b}(x)  
\label{b21}
\end{equation}
which  are  determined self-consistently later.
Taking a differentiation of  the  equation (\ref{b19}) we have 
a generalized  equation of BIPZ
\begin{equation}
P \int d\x \  {\r_0(\x) \over {x - \x}} = {V^{'}(x) \over {2\l}}
+ {1\over{2 \l}} \sum_{\a,\b =1}^K \w_{\a\b} W_{\a}^{'}(x) c_{\b} 
\end{equation} 
and it determines the stationary value $\r_0(x)$.  
Let us consider again a {\em one-cut solution, } 
i.e $\r_0(x)$ is equal to $zero$ 
if $x < a$ or $x > b$. Then the general solution 
is given by \cite{MI}:    
\begin{equation}
\r_0(x) = \int_a^b  d\x \ \hat{G}(x,\x)  
\bigg\{ V(\x)\ + 
\sum_{\a,\b =1}^K \w_{\a\b} W_{\a}(\x) c_{\b} \bigg\}  
+\ {1 \over \p} {1 \over \sqrt{(x-a)(b-x)}}      
\label{b30}
\end{equation}  
where $\hat{G}(x,y)$ is a differential operator
defined by   
\begin{equation}
 \hat{G}(x,y)\equiv {1\over{2\p^2 \l}}
 \left( \sqrt{(y-a)(b-y)\over{(x-a)(b-x)}} {P\over{y-x}}  \right)
 {\del \over{\del y}}.
 \label{b31}
\end{equation}
The coefficient of the homogeneous part (the second term) 
of $\rho_0(x)$ is 
determined so as to
satisfy the  constraint (\ref{b20}). 
\par
The constants     
$c_{\b}$'s are still unknown. In order to fix
them we plug equation (\ref{b30}) to the
equation (\ref{b21}) and  obtain a set of algebraic
equations for $c_{\b}$'s,  which can be
written in the following compact form:
\begin{equation}
\sum_{\g=1}^K \ \W_{\a\g} c_{\g} = O_{\a}   
\label{b70} 
\end{equation}
where 
\begin{eqnarray}
O_{\b} & \equiv & \int_a^b dx dy W_{\b}(x) \hat{G}(x,y) V(y)   
+ \int_a^b dx {W_{\b}(x) \over{\p \sqrt{(x-a)(b-x)}}}   
 \\
\W_{\b\g} & \equiv & \d_{\b\g} - 
\sum_{\a=1}^K \w_{\a\g}   
\int_a^b dx dy \ W_{\b}(x) \hat{G}(x,y) W_{\a}(y).  
\label{b60} 
\end{eqnarray} 
Assuming further that 
\begin{equation}
\det \ve \W \ve \neq 0
\label{b80}    
\end{equation}  
equation (\ref{b70}) can be inverted 
\begin{equation}
c_{\a} = \sum_{\b} (\W^{-1})_{\a\b} O_{\b}
\label{b82}   
\end{equation}  
to give the  solution for   
$\r_0(x)$:  
\begin{eqnarray}
\r_0(x) & = & \int_a^b  d\x \ \hat{G}(x,\x)  
\bigg\{ V(\x)\ + 
\sum_{\a,\b,\g =1}^K W_{\a}(\x) \w_{\a\b}
(\W^{-1})_{\b\g} O_{\g}  
\bigg\}  \nonumber \\ 
& + & \ {1 \over \p} {1 \over \sqrt{(x-a)(b-x)}}      
\label{b300}
\end{eqnarray} 
Now the only unknowns left are the end points of the cut
$[a,b]$. We can fix them     
by choosing boundary conditions $\r_0(a) = 0$ and
$\r_0(b) = 0$. These  
equations determine $a = a[V,W]$ and
$b = b[V,W]$ as functionals of $V(x)$ and $W(x)$'s.
Then equation (\ref{b300}) provides final expression 
for $\r_0(x)$.
\par
As in the previous section one can prove that
the solution (\ref{b300}) for
$\r_0(x)$ satisfies   
\begin{equation}
{\del \over{\del b}} \r_0(x) = 0. 
\label{b135} 
\end{equation} 
In order to prove it, notice that, 
from equation (\ref{b30}) and the boundary conditions 
$\r_0(a)=\r_0(b)=0$, we get
\begin{equation} 
{\del \over{\del b}} \r_0(x) = \int_a^b d\x \ 
\hat{G}(x,\x) \sum_{\a,\b=1}^K \ \w_{\a\b} 
W_{\a}(\x) {\del \over{\del b}} c_{\b}.   
\end{equation}  
All other contributions cancel
out due to the boundary conditions at the end points. 
Therefore taking derivative of $c_{\b}$ in equation
(\ref{b21}) we get a  set of homogeneous algebraic
equations, which can be written in the following
form
\begin{equation}
\sum_{\g=1}^K \ \W_{\b\g}{\del c_{\g} \over{\del b}} = 0. 
\label{b140} 
\end{equation}  
Since  
$\det \ve \W \ve \neq 0$, as previously assumed  
in  equation (\ref{b80}), we conclude
that 
\begin{equation}
{\del c_{\g} \over{\del b}} = 0 
\label{b150}
\end{equation}   
and  equation (\ref{b135}) is proved.
Similarly one can prove that
\begin{equation}
{\del \r_0(y) \over{\del a}} = 
{\del c_{\a} \over{\del a}} = 0. 
\label{b170} 
\end{equation}    
\par
Therefore upon variation of $\r_0(x)$ with respect to $V(x)$ only the 
manifest dependence on $V(x)$ in (\ref{b300}) remains.
For a connected two point correlation function we obtain: 
\begin{eqnarray}
&& N^2 \br \r(x) \r(y) \ke_c = -{\d \r_0(y) \over{\d V(x)}} = 
{1 \over{2\p^2 \l}} 
{\del \over{\del y}} \bigg\{ 
\sqrt{(y-a)(b-y)\over{(x-a)(b-x)}} {P\over{y-x}} 
\bigg\}    
\nonumber \\  
&& -\sum_{\a,\b=1}^K 
\int_a^b d\x \hat{G}(x,\x) W_{\a}(\x) \ \s_{\a\b} \ 
\int_a^b d\z \hat{G}(y,\z) W_{\b}(\z)  
\label{b400}   
\end{eqnarray}
where
\begin{eqnarray} 
\s_{\a\g}& \equiv& \sum_{\b=1}^K \ \w_{\a\b} 
(\W^{-1})_{\b\g}.   
\label{b500}
\end{eqnarray} 
Correlation function (\ref{b400})
is symmetric for $x$ and $y$ which follows from 
the symmetricity of the matrix $\omega$ and the definition 
of $\Omega$ (\ref{b60}).
\par
The correlation function (\ref{b400})
manifestly depends on $W_{\a}$'s and no longer
universal in a strong sense. 
However it is independent of $V(x)$ and also
the short distance behavior is dominated by the first term of 
the universal form,
which is what we call
{\em weak} form of universality.\footnote{
In the paper by B. Eynard and C. Kristjansen \cite{EK} they proved 
universality of correlation functions for $O(N)$ model on random 
lattice which can be written as one matrix model with an {\it infinite} 
sum of products of two traces. The universality might be recovered 
for their case since the {\it infinite} 
sum  may make the determinant eq. (\ref{b80})
divergent and $\sigma_{\alpha \gamma}$ in eq.(\ref{b500}) {\it zero}.
We would like to thank C. Kristjansen for calling our attention 
to their papers.}
It is  $O(N^0)$ and there is still strong suppression of
fluctuation though its amplitude is not universal.
And the second term of the correlation function
which breaks the universality is not inverse-proportional to
$\lambda$ generally. But 
if the second term in the definition of $\Omega$ eq.(\ref{b60})
is  much larger than $one$ 
the  correlation function (\ref{b400})
is again inverse-proportional to $\lambda$.

\par
The free energy $F[V]$ is written in the following form:
\begin{eqnarray}
F[V] &=& \int_{a[V,W]}^{b[V,W]} dx \  
V(\x) \r_0(\x) 
+ \sum_{\a,\b=1}^K \w_{\a\b} c_{\a} c_{\b} 
\nonumber \\ 
&-& \l \int_a^b dx dy \r_0(x) ln \ve x-y \ve \r_0(y)   
\label{b600}
\end{eqnarray}
where $c_{\alpha}$'s are defined in (\ref{b82}).
\par
Finally we  consider a case that
$\w_{12} = \w_{21} = 1$ 
and all other components equal to
{\em zero}. 
Using equation 
(\ref{b60}) and (\ref{b500}) 
the explicit form of the two-point correlation function is
\begin{eqnarray}
 && N^2 \br \r(x) \r(y) \ke_c =
{1 \over{2\p^2 \l}} 
{\del \over{\del y}} \bigg\{ 
\sqrt{(y-a)(b-y)\over{(x-a)(b-x)}} {P\over{y-x}} 
\bigg\} 
\nonumber \\  
 && - \int_a^b d\x \hat{G}(x,\x) W_{1}(\x) \ 
{(W_{2} \cdot \hat{G} \cdot W_{2}) 
\over{\det \ve \W \ve}}
\int_a^b d\z \hat{G}(y,\z) W_{1}(\z ) 
\nonumber \\  
 && -  \int_a^b d\x \hat{G}(x,\x) W_{1}(\x) \ 
{1 - (W_{1} \cdot \hat{G} \cdot W_{2}) 
\over{\det \ve \W \ve}}
\int_a^b d\z \hat{G}(y,\z) W_{2}(\z ) 
\nonumber \\  
&& -  \int_a^b d\x \hat{G}(x,\x) W_{2}(\x) \ 
{1 - (W_{2} \cdot \hat{G} \cdot W_{1}) 
\over{\det \ve \W \ve}}
\int_a^b d\z \hat{G}(y,\z) W_{1}(\z ) 
\nonumber \\  
 && -  \int_a^b d\x \hat{G}(x,\x) W_{2}(\x) \ 
{(W_{1} \cdot \hat{G} \cdot W_{1}) 
\over{\det \ve \W \ve}}
\int_a^b d\z \hat{G}(y,\z) W_{2}(\z ),
\label{b700}
\end{eqnarray} 
where
\begin{eqnarray}
&&\det \ve \W \ve \equiv 
( 1 - (W_2 \cdot \hat{G} \cdot W_1))
( 1 - (W_1 \cdot \hat{G} \cdot W_2) ) 
- (W_1 \cdot \hat{G} \cdot W_1) 
(W_2 \cdot \hat{G} \cdot W_2).
\label{b800} 
\nonumber \\
&&(W_{\a}\cdot \hat{G} \cdot W_{\b}) \equiv  
\int_a^b dx dy \ W_{\a}(x) \hat{G}(x,y) W_{\b}(y).  
\label{b801}  
\end{eqnarray}
A concrete example for $W_1=W_2=g_2 x^2 /2 +g_4 x^4/4$
is given in Appendix C.

\section{Discussion}

In this letter we studied density-density correlation functions 
for random matrix models with a generalized potential  containing
products of two traces 
$\tr W_1(M) \cdot \tr W_2(M)$ in addition to a single
trace $\tr V(M)$. We showed that the 
two point  function has no longer the universal 
form which depends on the potentials only through the end points.
This is against the argument by  Br\'{e}zin and Zee \cite{BZ93}.
They have considered a special ensemble
\begin{equation}
P (M) = {1 \over Z} e^{-N \tr V(M) - [\tr W(M)]^2 /2 }.
\end{equation}
They argued that the effect of the second term is just to 
renormalize the potential $V$ to $V+ \alpha_0 W$ where $\alpha_0$ is a constant
determined by $V$ and $W$ 
and  claimed that 
the universality for two point correlation function 
 still holds for the above  ensemble. 
Here we briefly review their argument and discuss
why it cannot generally hold for higher point correlation functions.
By introducing an auxiliary variable $\alpha$ the 
partition function is written as
\begin{eqnarray}
Z(V, W) &=& e^{- N^2 F[V, W]} = \int dM e^{-(N \tr V(M) + [\tr W(M)]^2/2 )}
\nonumber \\
&=& \int d \alpha \ e^{N^2 \alpha^2/2} e^{- N^2 F[V+\alpha W, 0]}
\end{eqnarray}
up to an irrelevant overall factor and the integral over $\alpha$ runs over the
imaginary axis.
In the large $N$ limit, the integral over $\alpha$ can be evaluated at the 
saddle point and 
\begin{equation}
F[V,W] = F[V+ \alpha_0 W, 0] - {\alpha_0^2 \over 2}
\end{equation}
where $\alpha_0$ is determined by 
\begin{equation}
\alpha_0 = { \partial \over \partial \alpha } F[V+ \alpha W, 0] |_{\alpha_0}.
\end{equation}
$\alpha_0$  depends  on the details of the potentials.
We can now obtain its density distribution function:
\begin{eqnarray}
\rho_0(x) 
&=& { \delta F[V, W] \over \delta V(x)}
\nonumber \\
&=& { \delta F[V+ \alpha_0 W, 0 ] \over \delta V(x)} |_{\alpha_0}
+ ({ \partial F[V+ \alpha_0 W, 0] \over \partial \alpha_0 } - \alpha_0)
{ \delta  \alpha_0 \over \delta V(x)}
\nonumber \\
&=& { \delta F[V+ \alpha_0 W, 0] \over \delta V(x)} |_{\alpha_0}.
\end{eqnarray}
This density distribution is 
the same as that of an ensemble with an
effective potential $\tr V_{eff} = \tr (V+ \alpha_0 W)$ 
as discussed in \cite{BZ93}. 
In order to obtain two point correlation function 
we then take variational derivative of the distribution
function with respect to the potential $V(y)$.
$\rho_0(x)$ depends on $V(y)$ not only explicitly but implicitly
through $\alpha_0$.
It becomes 
\begin{eqnarray}
 G(x, y) &=&  -{ \delta^2  F[V, W] 
               \over \delta V(x) \delta V(y)}
  = -{ \delta \rho_0(x)
               \over \delta V(y)}
\nonumber \\
        &=& - { \delta^2  F[V+ \alpha_0 W, 0 ] 
               \over \delta V(x) \delta V(y)}|_{\alpha_0}
            -  { \partial \rho_0(x) \over \partial \alpha_0 } 
    { \delta  \alpha_0 \over \delta V(y)}.
\end{eqnarray} 
The first term is the universal correlation function
for an effective potential $V_{eff}$ and 
actually independent of the details
of the potential.
But the second term, which is a product of a function of $x$ and that
of $y$, depends on the potential explicitly  and 
the universality is broken down.
(It is straightforward
to show that the above expression is equal to a special case 
of that obtained in section 3. )
In this simple case of potential,
 the extra term is factorized into functions of
$x$ and $y$ and 
we may say that there is still  universality in a weak sense.
As we discussed in the paper, if the potential term is more
complicated, other products of functions at $x$ and $y$ are added 
and this weak universality is also gradually broken.
\par
Our result can be also  generalized to potentials
containing products of more than two traces.
Two point correlation function is again given by the universal 
form plus a sum of products of a function of $x$ and that of $y$.
In this case we have to solve a non-linear equation to obtain
these functions.
\par
To conclude we have shown that the two point correlation function
is no longer universal in the strong sense and it depends on 
the details of the potentials. This implies that,
when we apply it to the random matrix models for transmission
matrix of a  quantum wire, 
conductance fluctuation is not
exactly universal. 
Amplitude of the fluctuation might depend
on the system size or disorder strength.
Also the variance of the conductance does not decrease exactly by 
a factor of two when magnetic field is applied.
\begin{center}
\begin{large}
Acknowledgments
\end{large}
\end{center}
We are deeply grateful to B. Sakita for his participation
in the early stages of this work, encouragement and many
valuable discussions. 
One of the authors (S. I.) would like to thank 
B. I. Halperin  and A. Zee for 
fruitful discussions. 
He  is grateful to the 
Institute for Advanced Study for its kind hospitality
and to Nishina Memorial Foundation for its financial support.
The other (An.K.) would like to thank V.P. Nair and G. Alexanian
for many valuable comments. This work was partially supported
by the National Science Foundation, grant number 
PHYS-9420615.     
\section*{Appendix A}
\setcounter{equation}{0}
\renewcommand{\theequation}{A.\arabic{equation}}

We prove
the following identity we have used in equation (\ref{a14}):
\begin{equation}
ln (\D^2) = \sum_{i\neq j}^N 
ln \ve x_i - x_j \ve 
=N^2 \int_{-\inf}^{+\inf} dx dy \r(x) ln \ve x-y \ve \r(y) 
+ const. 
\label{c10}
\end{equation}   
Regularizing $\d$-function in the definition of the
collective coordinate $\r(x)$ (see (\ref{a12})) as follows,       
\begin{equation}
\r(x) = \lim_{\e \to 0} {1\over{N}} 
\sum_{i=1}^N {(\e / \p) \over{(x-x_i)^2 + \e^2}}, 
\label{c20}
\end{equation} 
we obtain 
\begin{eqnarray}
&& N^2 \int_{-\inf}^{+\inf} dx dy 
\r(x) ln \ve x-y \ve \r(y) = 
\nonumber \\ 
&& = \lim_{\e \to 0} \sum_{i,j}^N \int_{-\inf}^{+\inf}dx dy   
{\e / \p \over{(x-x_i)^2 + \e^2}} \  ln \ve x-y \ve 
 \  {\e  / \p \over{(y-x_j)^2 + \e^2}}.
\label{c30}
\end{eqnarray} 
The sum can be separated into two pieces: $i \neq j$ and
$i = j$. The $i \neq j$ terms 
lead to the 
expression for the Van der Monde determinant.
The $i = j$ terms become, by setting 
$\x = x - x_i$ and $\z = y - x_i$,
\begin{eqnarray}
&& \lim_{\e \to 0} 
\sum_{i=1}^N \int_{-\inf}^{+\inf} dx dy  
{\e / \p \over{(x-x_i)^2 + \e^2}} \  ln \ve x-y \ve 
\  {\e  / \p \over{(y-x_i)^2 + \e^2}} = 
\nonumber \\
&& = \lim_{\e \to 0} 
\sum_{i=1}^N \int_{-\inf}^{+\inf} d\x d\z  
{\e / \p \over{\x^2 + \e^2}} \  ln \ve \x-\z \ve 
\  {\e  / \p \over{\z^2 + \e^2}} = const 
\label{c40}
\end{eqnarray} 
 and equation (\ref{c10}) is proved.  

\section*{Appendix B} 
\setcounter{equation}{0}
\renewcommand{\theequation}{B.\arabic{equation}}
Due to the following identity
\begin{equation}
f(y) \equiv {1\over{\p}}
\int_a^b dx {1 \over \sqrt{(x-a)(b-x)}} 
{P \over{y-x}} = 0,
\ \ \ \ \ \ \ \ \ \ a < y < b,  
\label{app10}
\end{equation}
the inhomogeneous term in   
$\rho_0(x)$ (equation (\ref{a30}) or equation (\ref{b30}) )
does not contribute to an integral (\ref{a20}).
This can be proved as follows.  
We first define  a  function 
\begin{equation}
F(z) = {1 \over \pi} \int_a^b  dx 
{1 \over \sqrt{(x-a)(b-x)}} {1 \over z-x}.
\label{app20}
\end{equation}
Then it follows   
\begin{equation}
f(y) = {1\over{2}} \bigg( 
F(y + i \epsilon) + F(y - i \e) 
\bigg),  \ \ \ \ \ \ \ \ \e \to 0.    
\label{app30}
\end{equation}
Choosing the square root to be positive on the upper 
side of the cut (and negative on the lower), 
the line integral (\ref{app20}) can be deformed to a contour 
integral along a path $C$ circling 
clockwise the cut between $a$ and $b$.
$z$ is outside of this contour.
Since the integrand  vanishes at
infinity the contour can be deformed smoothly to 
wind counterclockwise around $z$;
\begin{eqnarray}
F(z) &=& {1 \over 2 \pi} \oint_{C} dx {1 \over \sqrt{(x-a)(b-x)} }
                  {1 \over z - x}
\nonumber \\
&=& {1 \over 2 \pi} \oint_{z}  dx {1 \over \sqrt{(x-a)(b-x)} }
{1 \over z -x}
\nonumber \\
&=& i  {1 \over \sqrt{(z-a)(b-z)}}.
\end{eqnarray}
$F(z)$ has different signs on the upper
and lower sides of the cut and
we get
$f(x)=0$.  

\section*{Appendix C}
\setcounter{equation}{0}
\renewcommand{\theequation}{C.\arabic{equation}}
In this appendix we give an example for an ensemble with 
a potential 
$N \tr V(M) + (\tr W(M))^2$ where
\begin{equation}
W(x) = {g_2 x^2 \over 2} + {g_4 x^4 \over 4}.
\end{equation}
We assume that $V(x)$ is a symmetric potential and
$a=-b$. From eq. (\ref{b700}) we have 
\begin{eqnarray}
&&  N^2 \br \r(x) \r(y) \ke_c =
{1 \over{2\p^2 \l}} 
{\del \over{\del y}} \bigg\{ 
\sqrt{b^2 -y^2 \over{b^2-x^2}} {P\over{y-x}} 
\bigg\} 
\nonumber \\  
 && -  \int_{-b}^b d\x \hat{G}(x,\x) W(\x) \ 
{2 \over 1- 2 (W \cdot \hat{G} \cdot W)}
\int_{-b}^b d\z \hat{G}(y,\z) W(\z ) .
\end{eqnarray}
First it is straightforward to evaluate the integral
\begin{eqnarray}
&& \int_{-b}^b d\x \hat{G}(x,\x) W(\x) =
{1 \over 2 \pi^2 \lambda \sqrt{b^2 - x^2} }
\int_{-b}^b d\x \sqrt{b^2 - \x^2} {P \over{\x-x}} W'(\x)
\nonumber \\
&&= {1 \over 2 \pi^2 \lambda \sqrt{b^2 -x^2} }
\int_{-b}^b d\x \sqrt{b^2 - \x^2} 
\left[
g_2(1+x{P \over{\x - x}}) + g_4(\x^2 +
x \x + x^2 + x^3 {P \over{\x - x}})
\right]
\nonumber \\
&&= {1 \over 2 \pi^2 \lambda \sqrt{b^2 -x^2} }
\left[ g_2 ({\pi b^2 \over 2} - \pi x^2)
 + g_4 ({\pi  b^4 \over 8} + {\pi b^2x^2 \over 2} - {\pi x^4}) 
\right].
\end{eqnarray}
Then by using 
\begin{equation}
\int_{-b}^b d\z {\z^{2m} \over \sqrt{b^2 - \z^2} }=
 {(2m-1)!! \over 2m!!} \pi b^{2m},
\end{equation}
we can obtain 
\begin{eqnarray}
(W \cdot \hat{G} \cdot W) &=& 
-{1 \over \lambda} 
\left[ {g_2^2 b^4 \over 32}+{g_2 g_4 b^6 \over 32}
+{9 \over 1024} g_4^2 b^8 \right].
\end{eqnarray}
The end points $-b$ and $b$ are determined by the conditions
that the density distribution should vanish at end points.
The final answer for the connected density-density correlation
is in general not inversely proportional to $\lambda$.
This means that, when we apply matrix theory to the problem
of conductance fluctuation, the variance $(\delta G)^2$
does not decrease exactly by a factor of two.
However, if  
$|(W \cdot \hat{G} \cdot W)| >> 1$, 
we can {\em approximately}    
write the two point correlation function as
\begin{eqnarray}
&&  N^2 \br \r(x) \r(y) \ke_c =
{1 \over{2\p^2 \l}} 
{\del \over{\del y}} \bigg\{ 
\sqrt{b^2-y^2 \over{b^2 -x^2}} {P\over{y-x}} 
\bigg\} 
\nonumber \\  
 &+&
 {1 \over (2 \pi^2)^2 \lambda \sqrt{(b^2 -x^2)(b^2 - y^2)} }
\left[ g_2 ({\pi b^2 \over 2} - \pi x^2)
 + g_4 ({\pi  b^4 \over 8} + {\pi b^2 x^2 \over 2} - {\pi x^4}) 
\right] \times
\nonumber \\
&& \times
{1 \over {{1 \over{32}} g_2^2 b^4}+{{1\over{32}} g_2 g_4 b^6}
+{9 \over 1024} g_4^2   b^8 }
\left[ g_2 ({\pi b^2 \over 2} - \pi y^2)
 + g_4 ({\pi  b^4 \over 8} + {\pi b^2 y^2 \over 2} - {\pi y^4}) 
\right]
\nonumber \\
\end{eqnarray}
and it is again inversely proportional to $\lambda$. 


\end{document}